# Autonomous Energy Management system achieving piezoelectric energy harvesting in Wireless Sensors

[1]Sara Kassan, [2]Jaafar Gaber, [1]Pascal Lorenz

[1]Univ. Haute-Alsace UHA, 34 rue Grillenbreit, 68008 Colmar Cedex France, email: {sara.kassan, pascal.lorenz}@uha.fr

[2]FEMTO-ST Institute, Univ. Bourgogne Franche-Comté UBFC, Univ. Technology Belfort- Montbéliard UTBM, 13 rue E.T.Mieg, 90010 Belfort Cedex France, email: jaafar.gaber@utbm.fr

*Abstract*— Wireless Sensor Networks (WSNs) are extensively used in monitoring applications such as humidity and temperature sensing in smart buildings, industrial automation, and predicting crop health. Sensor nodes are deployed in remote places to sense the data information from the environment and to transmit the sensing data to the Base Station (BS). When a sensor is drained of energy, it can no longer achieve its role without a substituted source of energy. However, limited energy in a sensor's battery prevents the long-term process in such applications. In addition, replacing the sensors' batteries and redeploying the sensors is very expensive in terms of time and budget. To overcome the energy limitation without changing the size of sensors, researchers have proposed the use of energy harvesting to reload the rechargeable battery by power. Therefore, efficient power management is required to increase the benefits of having additional environmental energy. This paper presents a new self-management of energy based on Proportional Integral Derivative controller (PID) to tune the energy harvesting and Microprocessor Controller Unit (MCU) to control the sensor modes.

## I. INTRODUCTION

The possibility to avert replacing drained batteries is extremely important in Wireless Sensor Networks to avoid the high cost and time to replace batteries and redeploy them. Therefore, energy harvesting has attracted researchers and developers to recharge low power devices as Vivo-Nano-Robots, MEMS, Claytronics and WSN [1] [2] [3]. Energy harvesting is the method of extracting energy from the environment through different sources of energy. It is considered an emerging and reasonably mature technology to overcome the limited lifetime of battery-operated wearable devices and allows continuous recharging of the energy stored during use. The environmental energy for scavenging is mostly provided by ambient light (artificial lighting and solar lighting), radio frequency, thermal sources, kinetic and vibration sources [4] [5]. To harvest energy from vibration, there are different techniques such as electrostatic, electromagnetic, magnetostrictive and piezoelectric [6] [7] [8]. Table 1 compares the features of different type of vibration energy harvesting techniques.

*TABLE I: comparison of the different vibrational types of harvesting mechanisms*

| Technique type | Advantages | Disadvantages |
|---|---|---|
| Electromagnetic | -Strong coupling<br>-No smart materials | -Difficulty to be integrated by small Wireless Sensors<br>-Low voltage generated |
| Electrostatic | -Compatible with manufacturing Wireless Sensors<br>-Strong tensions generated | -Voltage source or load Required<br>-Mechanical constraints required<br>-Capacitive |
| Magnetostrictive | - High coupling<br>- No depolarization<br>- Flexibility | - Difficult small Wireless Sensor integration<br>- Fragility<br>- Nonlinear Effects<br>- Possible need for magnets polarization |
| Piezoelectric | Compact<br>-Compatible with integration small Wireless Sensors<br>-Strong tensions generated<br>-Strong coupling for single crystals | -Depolarization<br>-Fragility of the piezo layers<br>-High impedance |

Among them in this paper, we focus on the control of the low-level energy harvesting by the most prevalent technique piezoelectric vibration system for WSN. The diagram of a piezoelectric energy harvesting system is presented in Figure Fig. 1; it can be shortened into three essential components: piezoelectric devices, converters, and electrical energy storage. In addition to these components, we introduce two controllers one to tune the energy harvesting and the other to manage the energy in the sensor according to its activities needs. This new self-organization power system will be able to insure the prolongation of the WSN lifetime.

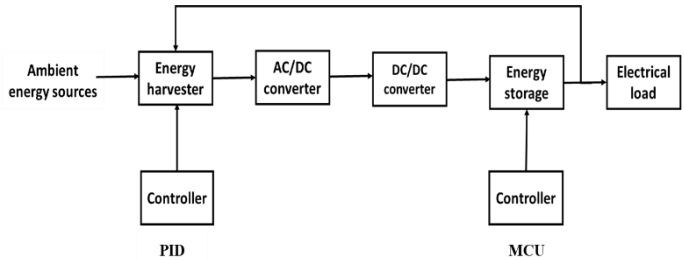

*Figure 1: Block diagram of energy harvesting model*

The paper is organized as follows: Section II provides a review with the different mechanisms to save energy in WSN. In section III, we present the global proposed model to manage the harvested energy in WSN. In section IV, we describe the piezoelectric harvesting energy model. In particular, we show how to adapt the PID controller to tune power in the vibration-based model in WSN and how to tune the residual energy in the rechargeable batteries according to the sensor's needs using MCU. In section V, we present the simulation results. Finally, we provide a conclusion of the paper in section VI.

## II. RELATED WORK

Prolonging WSN lifetime is one of the most critical challenges in WSNs. In the last few years, researchers have proposed different methods to increase the WSN lifetime via saving approaches. The main goal of these methods is saving energy, during the communication between sensors, without compensation for the energy dissipated via scavenging ambient energy. Specifically, we can classify energy methods in two types: energy saving mechanisms to decrease energy consumption and energy harvesting to recompense the energy consumed during sensors' activities.

Most of the proposed approaches in literature are saving energy-based mechanisms such as sensors deployment strategies, data-oriented techniques, topology control methods, and energy routing protocols. For example, in [9], the authors propose a statistical strategy for the nodes' deployment. The statistical node deployment strategy uses the Quasi-random method of low-discrepancy sequences to increase the lifetime and the coverage of the network.

The attentiveness to data-oriented techniques has increased recently. These techniques reduce the size of information via compression. Accordingly, data-oriented techniques are efficient to save energy during transmission and reception of this data information. In addition, the reduction of data is not only efficient in saving energy but also in saving more memory in the sensors [10] [11]. For example, in [11], the authors propose a Distributed Distortion-Rate Optimized Compressed Sensing (DQCS) method to compress data in WSN under a complexity-constrained encoding, which minimizes a weighted sum between the mean square error (MSE) signal reconstruction distortion and the average encoding rate.

A local lossless neighborhood indexing sequence (NIS) compression algorithm for data compression in WSN is proposed in [12]. The NIS approach dynamically assigns shorter length code information for each character in the input sequence, by exploiting the occurrence of neighboring bits, and every data packet is decompressed independently from others. Therefore, NIS algorithm helps to decrease the network load, which results in low data packet loss.

A topology control protocol based on learning automaton is proposed in [13]. The mechanism chooses the proper smallest transmission range of the node using the reinforcement signal produced by the learning automaton of neighbor sensor nodes. Consequently, the choice of transmission range affects the energy consumed by nodes and the overall network lifetime.

A hierarchical clustering protocol based on sensors location and energy consumption using non-cooperative game theory (GT) approach to extend the WSN lifetime is proposed in [14]. The GT permits a sensor to decide between two actions: to enter a game and transmit a message or to stay out of the game and harvest energy to charge its battery to reach the Nash Equilibrium (NE) solution for mixed strategies.

In [15], a Genetic algorithm-based Energy-efficient Clustering and Routing Approach (GECRA) is presented. This algorithm aims to calculate the total energy consumed by all sensor nodes where the algorithm encodes the clustering scheme and routing scheme together in the same chromosome.

It is worth noting that these research efforts aim mainly to reduce the energy consumption in the stage of wireless sensors deployment and during sensors activities, such as communication and sensing data, via energy saving mechanisms. Therefore, the WSN lifetime will be extended, but for a limited time. Instead of applying the energy saving methods, energy harvesting can be used to recharge the sensor's battery. The objective is to increase the power stored in the sensor's battery thereby extending the overall network lifetime depending on the availability of ambient energy harvesting resources. This article presents a new self-management of energy based on a Proportional Integral Derivative (PID) controller to tune the energy harvesting and a Microprocessor Controller Unit (MCU) to control the sensor modes.

## III. PROPOSED SELF POWER MANAGER SYSTEM MODEL

The target is an autonomous system to harvest energy spontaneously upon reaching a threshold of energy consumption. The system proposed is composed of two main blocks as designed in figure Fig. 2: The Battery Management Subsystem block that manages a rechargeable battery or a super-capacitor of the sensor and monitors its states via two (Management Controller Unit) MCU modes. In on mode, the sensor is active and can sense information from its environment, write/read data from memory, and communicate messages with other wireless sensors. During achieving activities, the sensor consumes energy until a threshold and the MCU turns to off mode. In the off mode, the block piezoelectric energy harvesting structure is triggered. This block is controlled via a control loop feedback using a Proportional Integral Derivate controller PID that scavenges ambient energy to recharge the sensor's battery.

In this paper, we assume that the vibration resource exists continuously. Moreover, the piezoelectric harvesting energy system is modeled by an equivalent Mass-Spring-Damper (MSD) model (discussed in the section below IV) with a PID controller that adapts the quantity of energy harvested by the sensor node to still alive and continue its processes. The harvesting energy will be tuned to reduce the error $e$ between the setpoint or reference energy $RE$.

When the error $e$ is bigger than a threshold $T$, the sensor node dissipates the majority of its residual energy and the microcontroller MCU switches to off mode and lets the sensors node harvest energy. Otherwise, the MCU is still in on mode and continues normally its activities. The energy consumption for one sensor depends on the energy consumed for its activities. The system block diagram is shown in Figure Fig. 2.

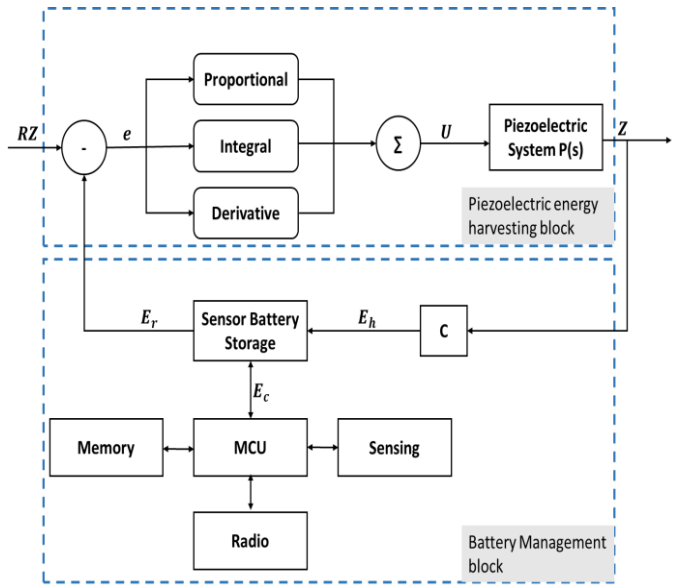

Figure 2: Architecture of the global system of wireless sensor composed of two main blocks one to manage energy consumed and residual energy in the battery and another one to harvest energy from the ambient source

In what follows, the model is formulated mathematically in terms of energy consumption by sensor nodes.
The energy error $e$ can be written as follows:

$$e(t) = RE - E_r(t) \qquad (1)$$

where the reference energy $RE$ is a fixed energy harvesting to charge the sensor's battery and $E_r$ is the current residual energy in the sensor's battery.

$$e(t) = \frac{1}{2}C\left(V_{ref}^2 - V_r(t)^2\right) \qquad (2)$$

where the reference voltage $V_{ref}$ is the desired fixed voltage and $V_r(t)$ is the current voltage for the sensor's battery in time $t$.

$$E_r(t) = E_h(t) + E_r(t-1) - E_c(t) \qquad (3)$$

It is required to find the total energy consumption of a node in the treatment of one data packet information. The total energy consumed by a wireless sensor node includes sensing, processing, communicating data information, switching radio model states and switching MCU modes.

- *Sensing energy consumption*

Sensing energy costs depends on the type of sensors. For example, the temperature sensors consumed less important energy than gas sensors. The sensor node can contain diverse sensors, and each one has its individual energy consumption attributes. Generally, the sensing energy consumption for a wireless sensor can be expressed as follows:

$$E_S = \alpha \times N \times V_{dc} \times I_{sens} \times T_{sens} \qquad (4)$$

where $I_{sens}$ is the required amount of current, and $T_{sens}$ is the duration to detect and collect $N$ bits data of information and $\alpha$ is the percentage to compress $N$ bits sensing data information.

- *Processing energy consumption*

The sensor consumes energy to read the data message and to write it in its memory. The processing energy consumption could be calculated by:

$$E_P = \frac{\alpha \times N \times V_{dc}}{8} \times (I_{Write} \times T_{Write} + I_{Read} \times T_{Read}) \qquad (5)$$

where $I_{Write}$ and $I_{Read}$ are the necessary amount current to write and read one byte data. $T_{Write}$ and $T_{Read}$ are the necessary duration to treat the $L(S_i)$ data information.

- *Communicating energy consumption*

The energy consumed to transmit and receive data messages is computed following the first-order wireless communication model for the radio hardware illustrated in Figure Fig. 3 [16].

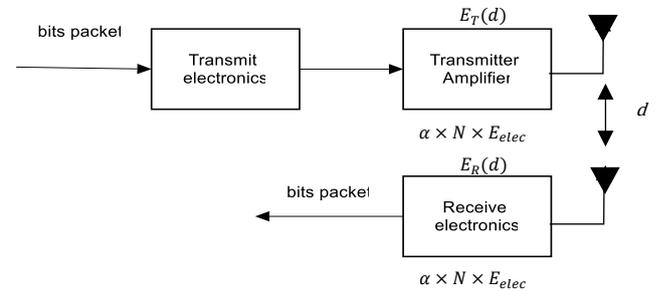

Figure 3: First order radio transceiver model

Transmitter expends energy to turn on the radio electronics and the power amplifier. The energy required to transmit $L$ bits data message is:

$$E_T = \begin{cases} \alpha \times N \times E_{elec} + \alpha \times N \times E_{fs} \times d^2 & \text{when } d < d_0 \\ \alpha \times N \times E_{elec} + \alpha \times N \times E_{mp} \times d^4 & \text{when } d > d_0 \end{cases} \quad (6)$$

where $E_{elec}$ represents the energy consumed to transmit or receive 1 bit message, the constants $E_{fs}$ and $E_{mp}$ depend on the transmitter amplifier model. $E_{fs}$ is for the free space model, $E_{mp}$ is for multipath model, $d$ is the distance between the transmitter and the receiver and $d_0$ is a threshold distance calculated as follows:

$$d_0 = \sqrt{E_{fs}/E_{mp}} \quad (7)$$

and the energy consumed by the radio to receive $L(S_i)$ bits data information is defined by:

$$E_R = \alpha \times N \times E_{elec} \quad (8)$$

- *Switching the microcontroller (MCU) mode energy consumption*

The sensor wastes energy by switching between the MCU modes. In this paper, we just take into consideration the on mode and the off mode. The energy cost for computational MCU mode can be expressed as:

$$E_{Switch-MCU} = V_{dc} \times (I_{Active} \times T_{Active} + I_{Sleep} \times T_{Sleep}) \quad (9)$$

IV. PIEZOELECTRIC ENERGY HARVESTING MODEL

In this section, the piezoelectric energy harvesting model is presented and then a PID controller is adapted to the model to tune the energy harvesting and the stability of the system.

For the piezoelectric energy harvesting model, the piezoelectric layers connected on a cantilever beam can be considered as a simple energy-harvesting device. The dimensions of piezoelectric, schematic and coordinate directions are shown in Figure Fig. 4. When the cantilever vibrates, at the first mode, the force which proceeds on the piezoelectric layers can be simplified to a 1-D model and regarded as a force $f$ acting on the lateral surface.

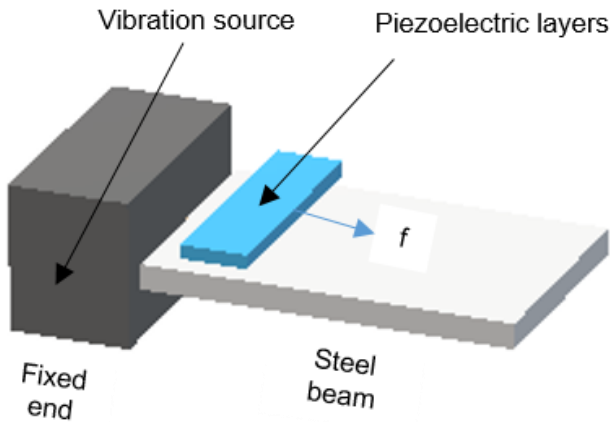

Figure 4: Schematic of piezoelectric harvesting structure

The mechanical characteristics of a piezoelectric heap can be represented by an equivalent Mass-Spring-Damper (MSD) model with one degree of freedom based on the proposed model in [17]. It consists of an equivalent mass $M_{pzt}$ linked with a spring that has a constant coefficient $K_{pzt}$ and a damper with a damping coefficient $D_{pzt}$ and input displacement of the frame $y(t)$.

The MSD model of the piezoelectric heap is presented in Figure Fig. 5.

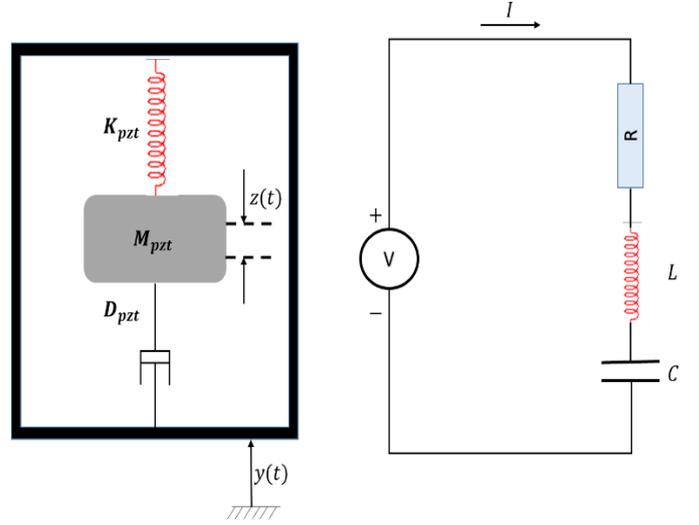

Figure 5: Equivalent MSD model

The displacement $z(t)$ of the mass relative to the frame is characterized by the differential equation (3), with the mass $M_{pzt}$, the spring constant; the equivalent short circuit stiffness; $K_{pzt}$, the damping coefficient $D_{pzt}$ and the input displacement of the frame $y(t)$. $V$ is the voltage across the load resistance, $C$ is the capacitance, $R$ is the resistance and $i$ is the electric current. The governing equation can be found by applying Newton's second law.

$$M_{pzt}\ddot{z}(t) + D_{pzt}\dot{z}(t) + K_{pzt}z(t) = f(t) \quad (10)$$

where

$$f(t) = -M_{pzt}\ddot{y} \quad (11)$$

We can present the equivalence mechanical equation of (13) by the electrical equation by applying Kirchhoff's law for an RLC circuit:

$$L\ddot{q}(t) + R\dot{q}(t) + \frac{1}{C}q(t) = V(t) \quad (12)$$

Where $L$ is the inductance, $R$ is the resistance, $C$ is the battery's capacitance and $q(t)$ is the quantity of electricity generated by the electromotive force voltage $V(t)$.

The electro-mechanical analogy of our system is represented as follows in Table 2:

TABLE II: Electro-mechanical analogy of piezoelectric energy harvesting system

| Mechanical system | Electrical system |
|---|---|
| Position $z(t)$ | Electrical charge $q(t)$ |
| Mass $M_{pzt}$ | Inductance $L$ |
| Damping coefficient $D_{pzt}$ | Resistance $R$ |
| Spring constant $K_{pzt}$ | Inverse of capacitance $C$ |
| External force $f(t)$ | Voltage $V(t)$ |

Applying the Laplace transform on equation (13) with zero initial conditions:

$$Z(s) = \mathcal{L}\{z(t)\} \text{ and } F(s) = \mathcal{L}\{f(t)\} \qquad (13)$$

$$M_{pzt}s^2 Z(s) + D_{pzt} s Z(s) + K_{pzt} Z(s) = F(s) \qquad (14)$$

The open-loop plant transfer function between the input vibration force; $F(s)$ ant the output displacement $Z(s)$ is given by:

$$P(s) = \frac{Z(s)}{F(s)} \qquad (15)$$

$$P(s) = \frac{1}{(M_{pzt}s^2 + D_{pzt}s + K_{pzt})} \qquad (16)$$

Figure Fig. 6 shows the displacement of the mass $M_{pzt}$ produced by an external vibration effort $F$. The DC gain (i.e, the amplitude ratio between the steady state response and the step input) is $1/K_{pzt}$, so $0.8116$ is the final value of the output for a unit step input. The rise time is about $0.4\ s$, the settling time is too long, about $6.562\ s$ and the overshoot is high, about $50.6432$.

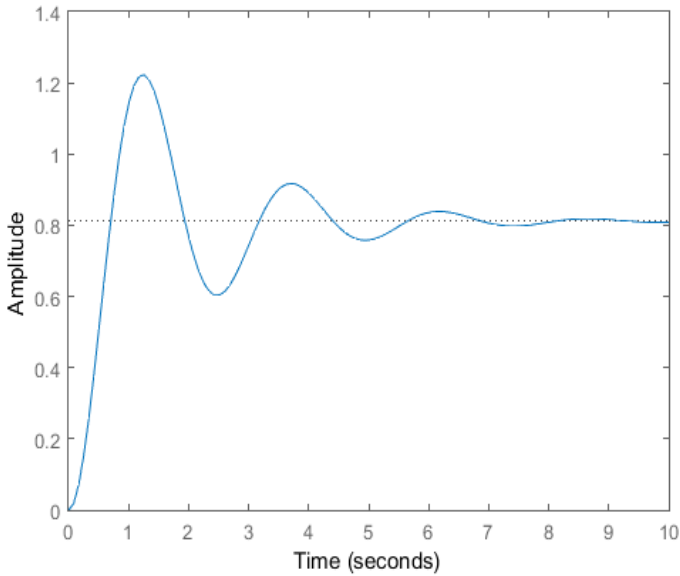

Figure 6: Transfer function between the input vibration force and the output displacement for an open loop without the PID controller

For the closed-loop, there are three transfer functions between the setpoint and the output under the PID controller and they are given by:
- The sensitivity function:

$$E(s) = \frac{1}{1 + P(s)U(s)} \qquad (17)$$

- The complementary sensitivity function:

$$T(s) = \frac{P(s)U(s)}{1 + P(s)U(s)} \qquad (18)$$

$$T(s) = \frac{(k_d s^2 + k_p s + k_i)}{(M_{pzt}s^3 + (D_{pzt}+k_d)s^2 + (K_{pzt}+k_p)s + k_i)} \qquad (19)$$

- The input sensitivity function:

$$I(s) = \frac{U(s)}{1 + P(s)U(s)} \qquad (20)$$

Proportional Integral Derivative (PID) controllers are extensively used to provide robustness and optimal performance for stable, unstable, and nonlinear processes [18]. It can be simply implementable in analog or digital form. Further, it supports tuning and online retuning based on the performance requirement of the process to be controlled. The output of a PID controller is calculated in the time domain from the response error and it can be expressed mathematically by the equation (21) as follows:

$$u(t) = k_d \frac{de(t)}{dt} + k_p e(t) + k_i \int_0^T e(t) \qquad (21)$$

where $u(t)$ and $e(t)$ are the control and the error signals respectively, $k_d$, $k_p$ and $k_i$ are the parameters to be tuned.
Taking the Laplace transform of equation (21), the transfer function for a PID controller can be expressed as follows:

$$U(s) = \frac{k_d s^2 + k_p s + k_i}{s} \qquad (22)$$

In this section, tuning the energy harvested by the piezoelectric system is indispensable to control the quantity of energy harvested, the stability of the system. Therefore, a Proportional Integral Derivative controller (PID) is adapted to the model represented above. The PID controller uses a control loop feedback mechanism to control the piezoelectric energy harvesting system. The parameters of the PID controller are the proportional gain $k_p$, the integral parameter $k_i$ and the derivative parameter $k_d$ that affect the control of the system.
The overall effects of controller parameters $k_p$, $k_i$ and $k_d$ on a closed-loop system are summarized in Table 3.

Improving proportional gain $k_p$ has the effect of equivalently increasing the control signal for the same level of error. The fact that the controller will drive harder for a given level of error lets the closed-loop system react more speedily, but also to overshoot more. In addition, increasing $k_p$ helps to decrease the steady-state error.

The addition of the integral parameter $k_i$ will continually increase over time to drive the steady-state error to attain zero. If there is a steady error, the integral response will slowly increase the control signal to the error down. However, the integral term can make the system more sluggish since when the error signal changes sign. It can take a while for the integral windup phenomenon when integral action saturates a controller without the controller driving the error signal toward zero.

The derivative controller parameter $k_d$ appends the ability of the control signal to become large, if the error begins sloping upward, even while the amplitude of the error is still quite small. This anticipation helps to add damping to the system, thereby decreasing overshoot. However, the addition of this parameter will not affect the steady-state error.

*TABLE III: The effects of PID controller parameters on a closed-loop system*

| PID parameters | Rise Time | Overshoot | Settling time | Steady-State error |
|---|---|---|---|---|
| $k_p$ | Decrease | Increase | Small changes | Decrease |
| $k_i$ | Decrease | Increase | Increase | Decrease |
| $k_d$ | Small changes | Decrease | Decrease | No change |

There are different methods to find the PID controller parameters that satisfy these goals. The fast rise time, the minimal overshoot, the minimal steady-state error and the minimal settling time. The gains of a PID controller can be found by trial and error method. Different methods have been proposed for setting the PID controller parameters. Some of these methods are based on characterizing the dynamic response of the dynamic system to be controlled with a first-order model or second-order model with a time delay. General methods for control design can be applied to PID control. A number of special methods that are tailor made for PID control have also been developed, these methods are often called tuning methods. The most well-known tuning methods are those that are specified by Ziegler and Nichols. The Ziegler-Nichols Oscillation Method, Ziegler-Nichol Process Reaction Method and Frequency Response method, is basic Self-Tuning methods. Oscillation method is based on system gain, in other words, system gain is redounded until the system makes oscillation, then PID parameters can be found from system response graphic. In our work, we apply Ziegler-Nichols heuristic tuning method to find the PID controller parameters [19] [20]. It consists to find the ultimate gain $K_u$ and the period $T_u$ by setting the Integral and Derivative gains to zero. The Proportional gain $G$ increases until it reaches the ultimate gain, at which the output displacement of the control loop has consistent oscillations. For a classic PID controller, we can find $k_p$, $k_i$ and $k_d$ by the relations as in Table 4:

*TABLE IV: The PID controller parameters according to Ziegler-Nichols tuning method on a closed-loop system*

| $k_p$ | $k_i$ | $k_d$ |
|---|---|---|
| $0.6K_u$ | $1.2\dfrac{K_u}{T_u}$ | $0.075K_uT_u$ |

The Flow Chart as shown in Figure Fig. 7 explains how to find the PID controller parameters based on Ziegler-Nichols heuristic tuning method.

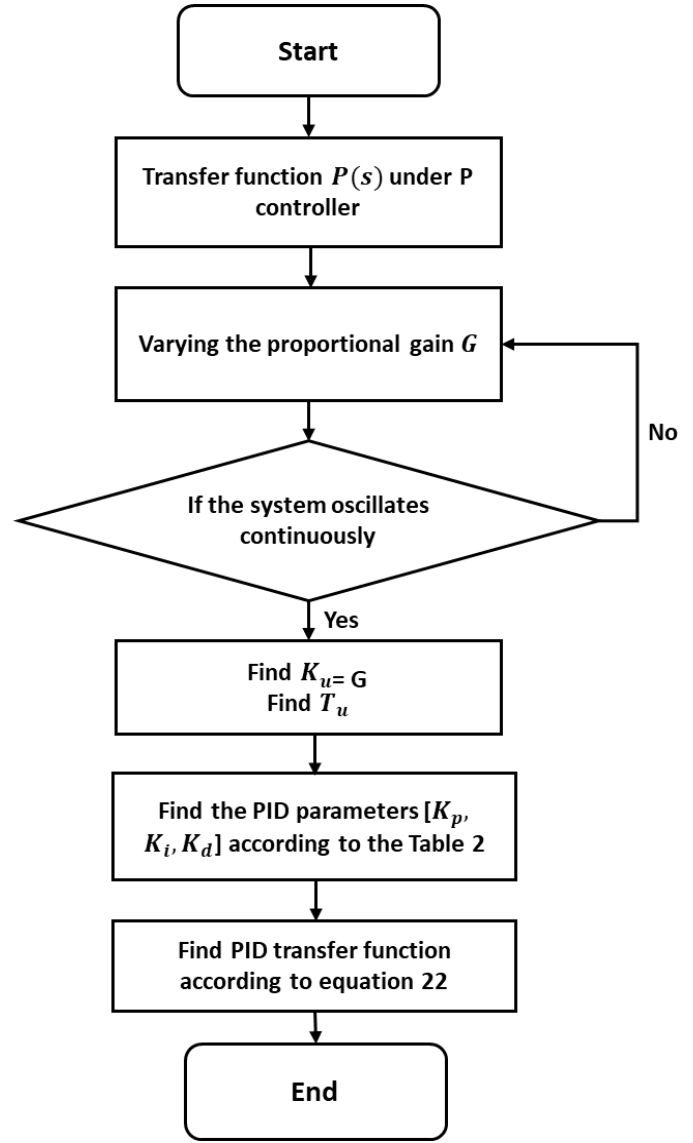

*Figure 7: Flowchart to find the transfer function for the PID controller based on Ziegler-Nichols heuristic tuning method*

Applying the Ziegler-Nichols heuristic tuning method, we find $K_u = 33.727$ and $T_u = 3.90176$. The transfer function for the PID controller can be found from eq. 2 and can be rewritten as follows:

$$u(s) = \frac{9.8699s^2 + 20.2366s + 10.3729}{s} \quad (23)$$

Applying Routh-Hurwitz criterion, we can verify if the system in a closed loop converges, stable or not [21].
The following structure of Routh-Hurwitz criterion matrix for the complementary sensitivity transfer function is shown as follows:

$$\begin{pmatrix} M_{pzt} & (K_{pzt} + k_p) \\ (D_{pzt} + k_d) & k_i \\ \frac{\left((D_{pzt} + k_d) \times (K_{pzt} + k_p)\right) - \left(k_i \times M_{pzt}\right)}{(D_{pzt} + k_d)} & 0 \\ k_i & 0 \end{pmatrix} \quad (24)$$

All the values of column 1 of the table are positive. There is no sign changed. Therefore, the system controlled by pour PID controller is stable for the closed loop.

Figure Fig. 8 shows the displacement of the mass $M_{pzt}$ produced by an external vibration effort $F$ controlled by our PID controller. This controller is designed based on Ziegler-Nichols tuning method. The new system has a fast rise time $0.0311\ s$, it reduces also the settling time $0.0509\ s$, and minimize the overshoot $1.1887$.

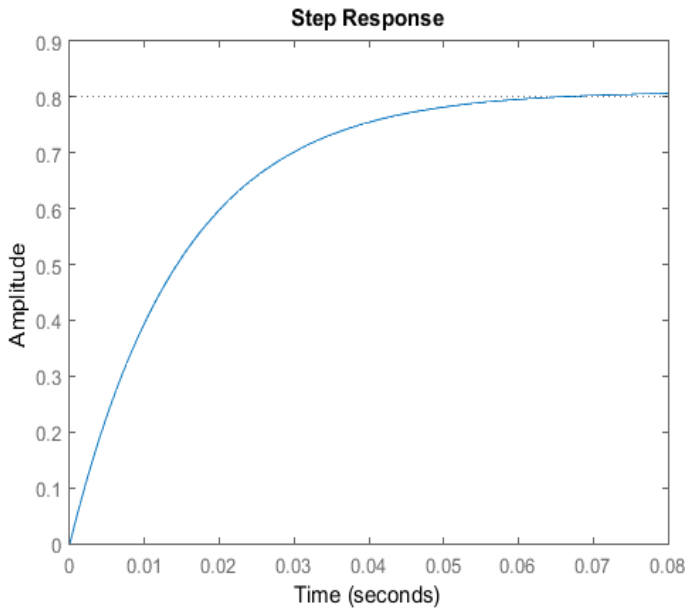

*Figure 8: Transfer function between the input vibration force and the output displacement for a closed-loop with an adaptive PID controller based on Ziegler-Nichols tuning method.*

The response of the MSD to a unit step input without a PID controller is shown in Figure Fig. 6, where it is observed that the MSD system without a PID controller is oscillating in nature. It is clear that the settling time is about $6.562\ s$ and the peak overshoot is about $50.6432\ p.u.$. The rise time is about $0.4\ s$ and the settling time is too long, about $6.562\ s$. In addition, the terminal voltage reaches a steady value of $0.8116\ p.u.$ at about $9{,}2\ s$, which is not acceptable. We can enhance the MSD system by using a controller. The using of a proportional controller can reduce the rise time of a response, but the steady-state error cannot be removed. An integral controller can improve the steady-state performance, but it can also badly affect the transient response. A derivative controller increases the performance of the transient response by reducing overshoot, thereby improving the stability margin for the system. The behavior of the MSD system can be improved by including a PID controller as shown in Figure Fig. 8.

## V. SIMULATION RESULTS AND PERFORMANCE

The values of the hardware parameters used in our simulations are those of Mica2 Motes. All parameters used in our energy model are listed in Table 5. We also indicate the references where these values originated. Our results are based on existing device parameters. Moreover, that is why they can reflect a real energy model for a type of wireless sensors.

*TABLE V: Simulation parameters*

| Parameter | value |
|---|---|
| The supply voltage to the sensor $V_{dc}$ (V) | 2.7 |
| Reference Energy RE (J) | 0.2 |
| Threshold energy | 0.1 |
| Sensing time for one bit of data $T_{Sens}$ (ms) | 0.5 |
| Sensing current $I_{Sens}$ (mA) | 25 |
| Writing current $I_{Write}$ (mA) | 18.4 |
| Writing time for one bit $T_{Write}$ (ms) | 12.9 |
| Reading current $I_{Read}$ (mA) | 6.2 |
| Reading time for one bit $T_{Read}$ (µs) | 565 |
| On mode for the MicroController Unit (MCU) current $I_{Active}$ (mA) | 8 |
| On mode for the MicroController Unit (MCU) time $T_{Active}$ (ms) | 1 |
| Off mode for the MicroController Unit (MCU) current $I_{Sleep}$ (µA) | 1 |
| Off mode for the MicroController Unit (MCU) time $T_{Sleep}$ (ms) | 299 |
| Initial energy (J) $E_0$ | $0.5\ \alpha$ |
| Energy dissipation: electronics $E_{elec}$ (nJ/bit) | 50 |
| parameters of amplifier energy consumption $Emp$ (pJ/bit/m4) and $Efs$ (pJ/bit/m2) | 0.0013 and 10 |
| Data aggregation energy (J) | $5 \times 10^{-12}$ |
| Size of the data packet (bits) $N$ | 4000 |
| Compression percentage (%) $\alpha$ | 20 |
| Mass $M_{pzt}$ (g) | 182 |
| The damping ratio of the structure $D_{pzt}$ (N.m$^{-1}$.s$^{-1}$) | 0.2 |

| Parameter | value |
|---|---|
| The supply voltage to the sensor $V_{dc}$ (V) | 2.7 |
| Spring stiffness constant $K_{pzt}$ (N.m$^{-1}$) | 0.12320 |
| Proportional controller parameter $K_p$ | 20.2366 |
| Integral controller parameter $K_i$ | 10.3729 |
| Derivative controller parameter $K_d$ | 9.8699 |

Figure Fig. 9 shows the residual energy in the sensor's battery. When the battery is out of energy, the MCU switches to off mode. During the off mode, the battery is charged with a vibration harvesting energy without the adaptive PID controller. The battery is recharged to $0.1245\ Joules$ and cannot reach more than this value. Therefore, there is a need to include an adaptive PID controller to the system that can enhance the stability and the performance of the system.

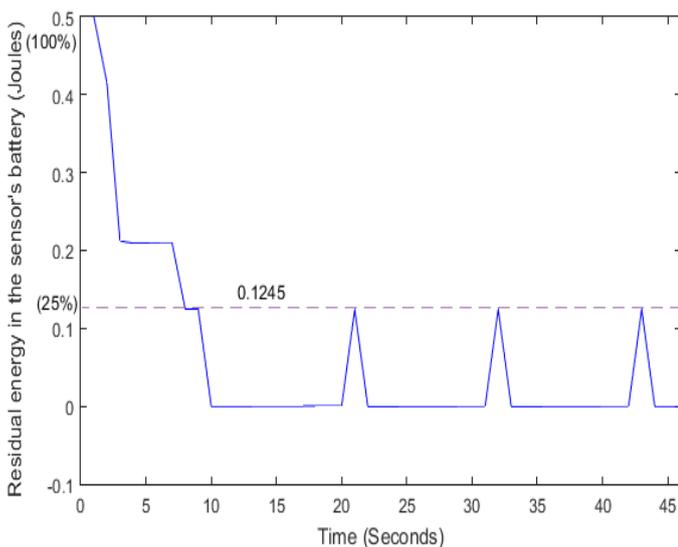

*Figure 9: Self-power manager behavior without the adaptive PID controller*

Figure Fig. 10 shows the residual energy in the sensor's battery. When the residual energy is less than a given threshold value (e.g. 20%,) of the initial energy $0.5\ Joules$ with $\alpha = 1$, the MCU switches to off mode. During the off mode, the battery is charged with a vibration harvesting energy controlled by the adaptive PID controller. The battery is recharged to a fixed threshold $0.3249\ Joules$.
The results illustrate that the system is still cyclical stable even if the energy consumption depends on different sensor's activities and it is not the same for each time period.

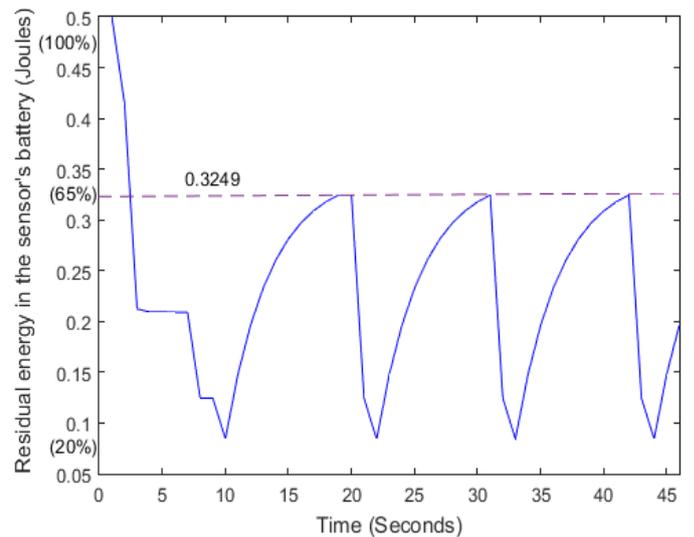

*Figure 10: The self-power manager controller organizes its wake-up mode to do its activities and its off mode to harvest energy controlled by the adaptive PID controller depending on its residual energy*

The self-power manager system including the PID controller can recharge 65% of the battery as shown in Figure Fig. 9 while the system without the adaptive PID controller cannot charge more than 25% at the same period of time as shown in Figure fig. 9. Therefore, the self-power manager system must be tuned by an adaptive PID controller, to recharge the battery spontaneously (according to a critical threshold value), and stabilize the system whatever are the sensor's activities.

## VI. CONCLUSION

An adaptive energy manager is essential to harvest energy in WSNs. Its core goal is to control the energy harvesting to maximize the lifetime of the sensor nodes and give the sensors the power to achieve their activities. In this paper, autonomous power management with a PID controller provides practical adaptation to the harvested and consumed energy for a sensor node. Moreover, an adaptive controller for the MCU modes, to extend a sensor lifetime depending on the existing vibration in the environment, is presented. Future work will cover a combination between simulation results and real experimental validation of the proposed autonomous energy management system. In addition, we will focus our work on the integration of some existing management energy efficient methods in our self-organized power manager system to save more energy during communications and to reduce the rapid consumption of residual energy in the sensor's battery and from the environmental energy resources. Applying energy efficient protocols for sensors' communications with the autonomous energy management system will increase the overall WSN lifetime.

ACKNOWLEDGMENT

The authors would like to thank the editors and reviewers for their valuable comments on earlier versions of this paper.